\documentclass[aps,amsmath,amssymb,twocolumn]{revtex4-1}

\usepackage{graphicx}
\usepackage{dcolumn}
\usepackage{bm}
\usepackage{color}

\begin{document}

\title{Proximity Effect between two Superconductors Spatially Resolved by Scanning Tunneling Spectroscopy}

\author{V. Cherkez$^{1,2}$}
\author{J. C. Cuevas$^3$}
\email{juancarlos.cuevas@uam.es}
\author{C. Brun$^{1,2}$}
\email{christophe.brun@insp.upmc.fr}
\author{T. Cren$^{1,2}$}
\author{G. M\'{e}nard$^{1,2}$}
\author{F. Debontridder$^{1,2}$}
\author{V. Stolyarov$^{1,2,4}$}
\author{D. Roditchev$^{1,5}$}

\affiliation{$^1$Sorbonne Universit\'{e}s, UPMC Univ Paris 06, UMR 7588, Institut des Nanosciences de Paris, 
F-75005, Paris, France}
\affiliation{$^2$CNRS, UMR 7588, Institut des Nanosciences de Paris, F-75005, Paris, France}
\affiliation{$^3$Departamento de F\'{\i}sica Te\'orica de la Materia Condensada and Condensed Matter 
Physics Center (IFIMAC), Universidad Aut\'onoma de Madrid, E-28049 Madrid, Spain}
\affiliation{$^4$ Moscow Institute of Physics and Technology, 141700 Dolgoprudny, Russia}
\affiliation{$^5$LPEM, ESPCI ParisTech-UPMC, CNRS-UMR 8213, 10 rue Vauquelin, 75005 Paris, France}

\date{\today}

\begin{abstract}
We present a combined experimental and theoretical study of the proximity effect in an atomic-scale
controlled junction between two different superconductors. Elaborated on a Si(111) surface, the
junction comprises a Pb nanocrystal with an energy gap $\Delta_1=1.2$~meV, connected to a crystalline
atomic monolayer of lead with $\Delta_2=0.23$~meV. Using \textit{in situ} scanning tunneling spectroscopy
we probe the local density of states of this hybrid system both in space and in energy, at temperatures
below and above the critical temperature of the superconducting monolayer. Direct and inverse proximity
effects are revealed with high resolution. Our observations are precisely explained with the help of
a self-consistent solution of the Usadel equations. In particular, our results demonstrate that in the 
vicinity of the Pb islands, the Pb monolayer locally develops a finite proximity-induced superconducting 
order parameter, well above its own bulk critical temperature. This leads to a giant proximity effect where
the superconducting correlations penetrate inside the monolayer a distance much larger than in a
non-superconducting metal.
\end{abstract}


\maketitle

\section{Introduction}

If a normal metal ($N$) is in good electrical contact with a superconductor ($S$), Cooper pairs can
leak from $S$ to $N$, modifying the properties of the metal. This phenomenon, known as \emph{proximity
effect}, was intensively studied in the 1960's \cite{deGennes1964,Deutscher1969} and there has been
a renewed interest in the last two decades due to the possibility to study this effect at much smaller
length and energy scales \cite{Pannetier2000}. When a Cooper pair penetrates into a normal metal, via
an Andreev reflection \cite{Andreev1964}, it becomes a pair of time-reversed electron states that
propagate coherently over a distance $L_C$, which in diffusive metals is given by $L_C = \mbox{min}
\{\sqrt{\hbar D/E}, L_{\phi}\}$, where $D$ is the diffusion constant, $E$ is the energy of the electron
states (with respect to the Fermi energy), and $L_{\phi}$ is the phase-coherence length in $N$. This
Cooper pair leakage modifies the local density of states (DOS) of the normal metal over a distance
$L_C$ from the $S$-$N$ interface. Such a modification has been spatially resolved in recent years with
the help of tunneling probes \cite{Gueron1996,Meschke2011} and with Scanning Tunneling Microscopy/Spectroscopy
(STM/STS) techniques applied to mesoscopic systems \cite{Chapelier2001,Moussy2001,Escoffier2004,leSueur2008,Wolz2011}.
Very recently, the considerable progress in the controlled growth of atomically clean materials under
ultrahigh vacuum conditions has made it possible to probe the proximity effect with high spatial and energy
resolution in \emph{in situ} STM/STS experiments \cite{Kim2012,Serrier-Garcia2013}.

The proximity effect is not exclusive of $S$-$N$ systems. If a superconductor $S_1$, with a
critical temperature $T_{C1}$ and energy gap $\Delta_1$, is brought into contact with another
superconductor $S_2$ with a lower critical temperature $T_{C2} < T_{C1}$ and
energy gap  $\Delta_2 < \Delta_1$, the local DOS of both superconductors near the interface is expected
to be modified. At low enough temperature $T < T_{C2}$ this modification should be significant in the
energy interval $|E| \in [\Delta_{1}, \Delta_{2}]$ and may occur in each electrode over a distance
$\mbox{min}\{\sqrt{\hbar D/\Delta_i}, L_{\phi_i}\}$ from the interface. Moreover, in the temperature
range $T_{C2} < T < T_{C1}$ one expects the proximity effect to induce a finite local order parameter
in a formally non-superconducting $S_2$, owing to a non-zero attractive pairing interaction $\lambda_2$
existing in $S_2$. Such a mechanism should result in a proximity-induced interface superconductivity.
These remarkable effects were first discussed qualitatively by de Gennes and co-workers in the 1960's
\cite{deGennes1964,Deutscher1969}, but to the best of our knowledge no experiment has ever been reported
in which this peculiar $S_1$-$S_2$ proximity effect could be spatially resolved.

In this work we present a STM/STS study of the proximity effect in a lateral $S_1$-$S_2$ junction
with a very high spatial and energy resolution, for temperatures well below and above $T_{C2}$. The
junction was elaborated \emph{in situ} in ultrahigh vacuum on Si(111). The $S_1$ electrode is formed
by a single nanocrystal of Pb ($T_{C1} \approx 6.2$~K, $\Delta_1 = 1.2$~meV). $S_2$ consists of a
single atomic layer of Pb reconstructed on Si(111) to form the so-called striped incommensurate phase,
a superconductor with $T_{C2} \approx 1.8$~K and $\Delta_2 = 0.23$~meV. For temperatures well below
$T_{C2}$ we observe a pronounced modification of the local tunneling conductance spectra in $S_2$ over
a distance $\sim 100$~nm from the $S_1$-$S_2$ interface. Above $T_{C2}$ but below $T_{C1}$, the tunneling
spectra in $S_2$ exhibit an induced gap which extends to distances anomalously large for a normal metal.
Our experimental observations are explained with the help of an one-dimensional model based on the Usadel
equations, where the order parameter is evaluated self-consistently. Importantly, our combined experimental
and theoretical study furnishes strong evidence that the long-range proximity effect observed in $S_2$
above $T_{C2}$ is a direct consequence of the appearance of proximity-induced interface superconductivity
in the atomic monolayer.

The paper is organized as follows. Section \ref{sec-II} describes the fabrication of
the atomic-scale controlled junctions between two different superconductors and present
the results for the tunneling spectra in the low temperature regime where both electrodes
are in the superconducting state. Section \ref{sec-III} is devoted to the discussion of the theoretical
model based on the Usadel theory that is used to describe all our experimental results. We present and
discuss in section \ref{sec-IV} the results for the tunneling spectra in the temperature regime where
the Pb monolayer is in its normal state. Then, we briefly discuss the inverse proximity effect in the
Pb islands and summarize our main conclusions in section \ref{sec-V}.

\section{Proximity effect between two superconductors: Experimental setup and low temperature results} \label{sec-II}

The fabrication of a $S_1$-$S_2$ junction was carried out \emph{in situ} as we proceed to explain. First, a 
$7 \times 7$ reconstructed \emph{n}-Si(111) surface (heavily doped) was prepared by direct current heating at 
1200$^{\circ}$C in ultrahigh vacuum. Subsequently, 1.65 monolayer of Pb was evaporated on the Si(111)-$7\times7$ kept 
at room temperature, using an electron beam evaporator calibrated with a quartz micro-balance. The $\surd7 
\times \surd3$ reconstructed Pb monolayer was formed by annealing at 230$^{\circ}$C for 30 minutes \cite{Kumpf2000}. 
The slightly denser striped incommensurate phase was then formed by adding 0.2 monolayer (ML) of Pb onto the 
$\surd7 \times \surd3$-Pb/Si(111) held at room temperature \cite{Seehofer1995,Horikoshi1999}. This resulted in 
a slight extra amount of lead atoms (0.07~ML) with respect to the nominal reported coverage of the striped 
incommensurate phase (1.33~ML).

\begin{figure}[t]
\begin{center} \includegraphics[width=\columnwidth,clip]{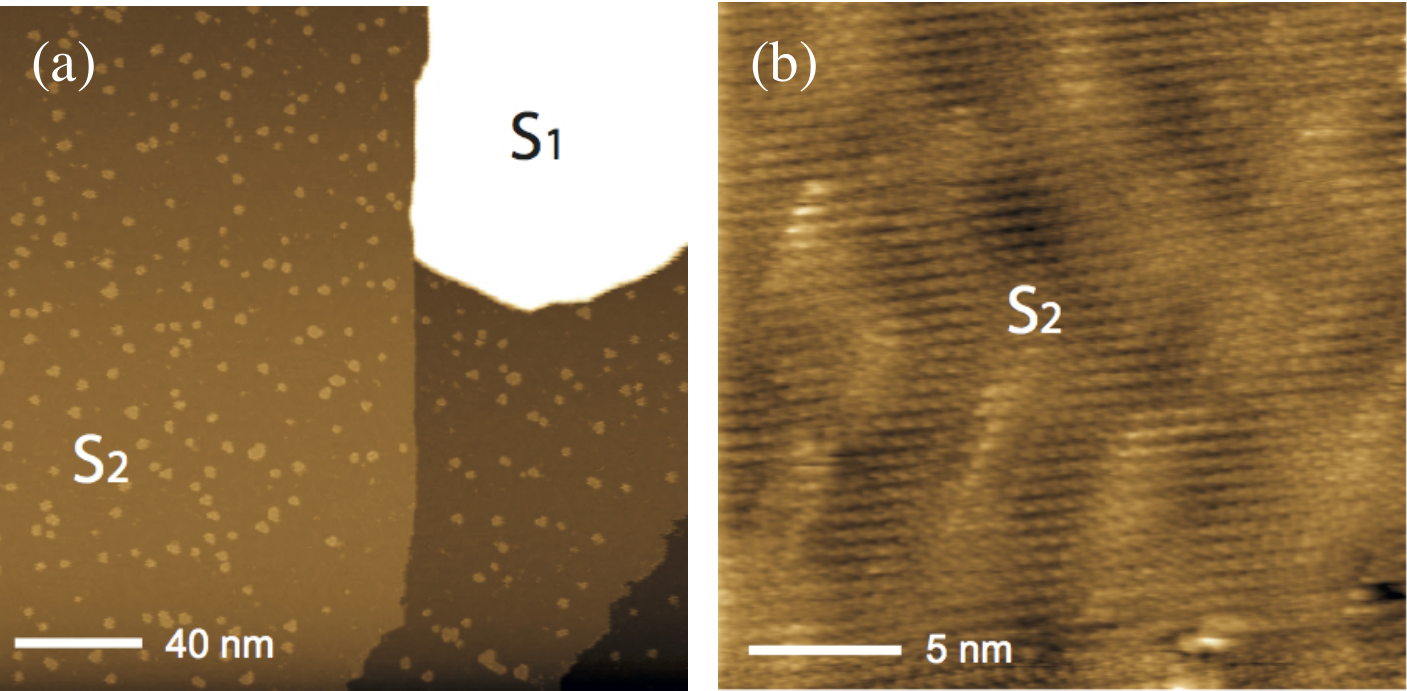} \end{center}
\caption{(a) Topographical STM image of a 7~ML high Pb island, denoted as $S_1$, surrounded by the striped
incommensurate (SIC) Pb monolayer, denoted as $S_2$. Small Pb clusters of few nanometers size and 1~ML height
are visible on the SIC monolayer. These very small clusters result from the extra Pb atoms deposited on the
surface in order to form few large islands such as $S_1$. (b) Topographical STM image showing a smaller scale
region representing the atomic superstructure of the SIC monolayer observed everywhere on the surface between
the Pb clusters. The images were taken at $V_{\rm bias}=$ -1.0~V and $I=$ 35~pA.}
\label{topo_proxy}
\end{figure}

This procedure allowed us to grow a very small density of Pb islands of size larger than 100~nm and height of
7~ML, such as the one denoted by $S_1$ in the STM topographic image shown in Fig.~\ref{topo_proxy}(a). These islands
have a critical temperature and an energy gap slightly smaller than the bulk Pb values, here $T_{C1} \approx 6.2$~K 
and $\Delta_1 \approx 1.2$~meV \cite{Eom2006,Nishio2008,Cren2009,Brun2009}. The island $S_1$ shown in 
Fig.~\ref{topo_proxy}(a) is in direct electrical contact through peripheral atoms with the striped incommensurate 
(SIC) monolayer denoted $S_2$, reported to be a superconductor with $T_{C2} \approx 1.8$~K and $\Delta_2
\approx 0.3$~meV \cite{Zhang2010,Yamada2013}. It has been shown by various surface techniques that the Pb
islands lay directly on top of the Si(111)-$7\times7$ substrate \cite{Feng2004}. In addition, one can see in
Fig.~\ref{topo_proxy}(a) the formation of very tiny 1~ML high nanoprotrusions or clusters on top of the striped 
incommensurate phase (size less than 5~nm), also resulting from the deposition of the extra 0.07~ML of Pb. As 
presented in Fig.~\ref{topo_proxy}(b), the atomic superstructure of the SIC was observed everywhere on the surface 
between the overlying Pb islands or clusters. At each step, the sample structure was controlled in both real and 
reciprocal space by STM and Low Energy Electron Diffraction. The STS measurements were performed \emph{in situ} 
with a homemade apparatus, at a base temperature of 320~mK and in ultrahigh vacuum $P<4.0$ $10^{-11}$~mbar 
\cite{Cren2009,Cren2011}. Mechanically sharpened Pt-Ir tips were used. The tunneling conductance curves $dI(V)/dV$ 
were obtained from numerical derivatives of the raw $I(V)$ experimental data.

\begin{figure*}[t]
\includegraphics*[width=\textwidth,clip]{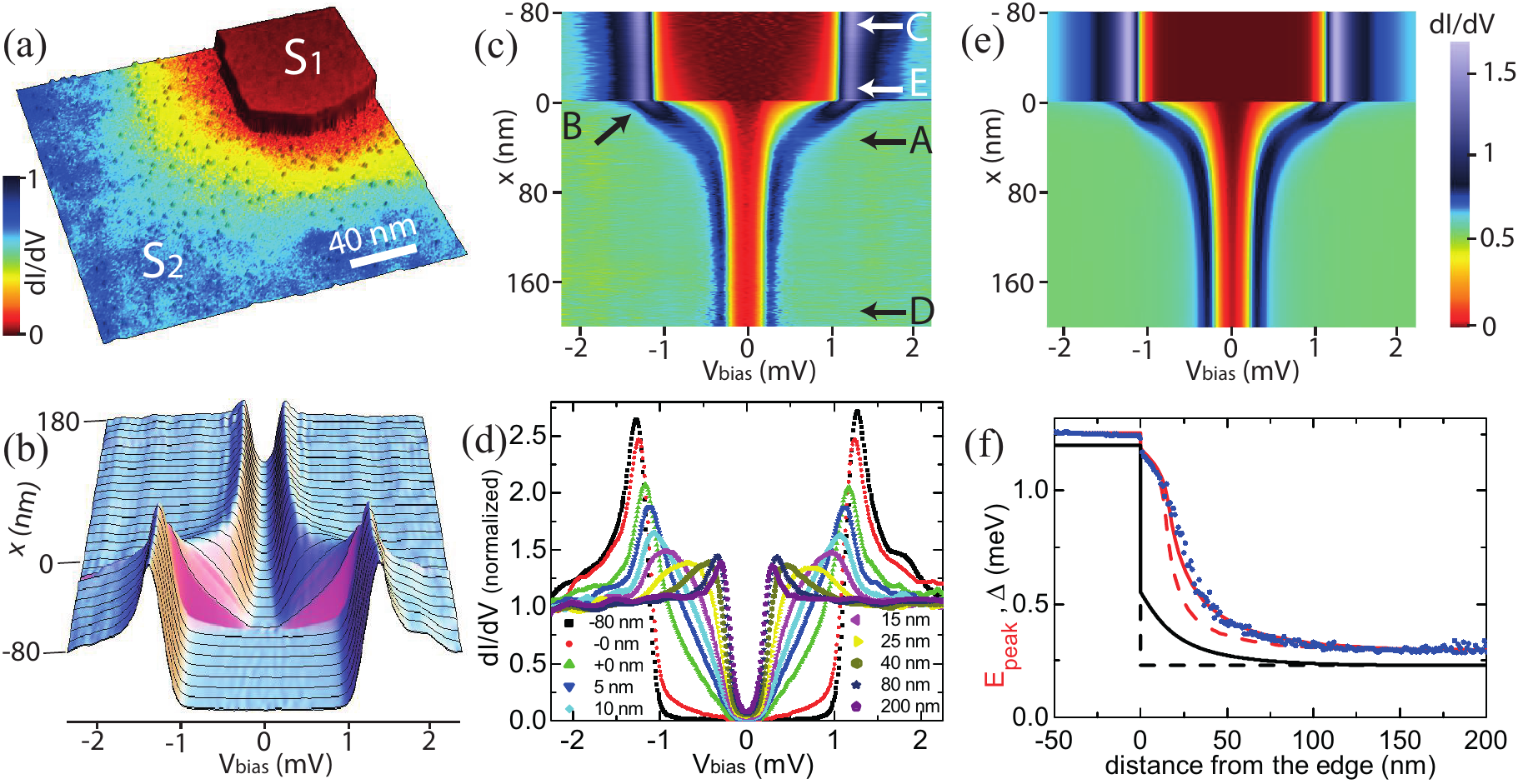}
\caption{$S_1$-$S_2$ proximity effect at 0.3~K. (a) Topographic STM image of the sample showing the
Pb nanoisland $S_1$ connected to the striped incommensurate Pb monolayer $S_2$. Superposed color-coded
spectroscopy map at $V_{bias}=-0.2$~mV allows visualizing the proximity effect. 256$\times$256 spectra
were measured in the STS map. (b) Spatial and energy evolution of the experimental tunneling conductance
spectra, $dI/dV(V,x)$, across the junction (3D-view). One spectrum is plotted every 1~nm and highlighted 
by a black line every 10~nm. (c) Color-coded
experimental $dI/dV(V,x)$ spectra across the interface. One spectrum is plotted every nanometer. (d)
Selected local tunneling spectra (dots). $-0$~nm is the last spectrum measured on the top flat part of
the island before the edge. $+0$~nm is the first spectrum measured on the SIC monolayer. The distance
between the $+0$~nm and $-0$~nm spectra is about 1 nanometer (e) Color-coded computed $dI/dV(V,x)$ across
the interface. (f) Spatial evolution of the energy of the peak maximum $E_{peak}(x)$ across the interface.
The experimental results (symbols) are nicely reproduced by self-consistent calculation of the order parameter
(red solid line), while the red dashed line corresponds to the non-self-consistent result. The evolution of
the order parameter is shown by black lines: self-consistent (solid) and non-self-consistent (dashed).}
\label{fig_exp_lowT}
\end{figure*}

\begin{figure*}[t]
\includegraphics*[width=\textwidth,clip]{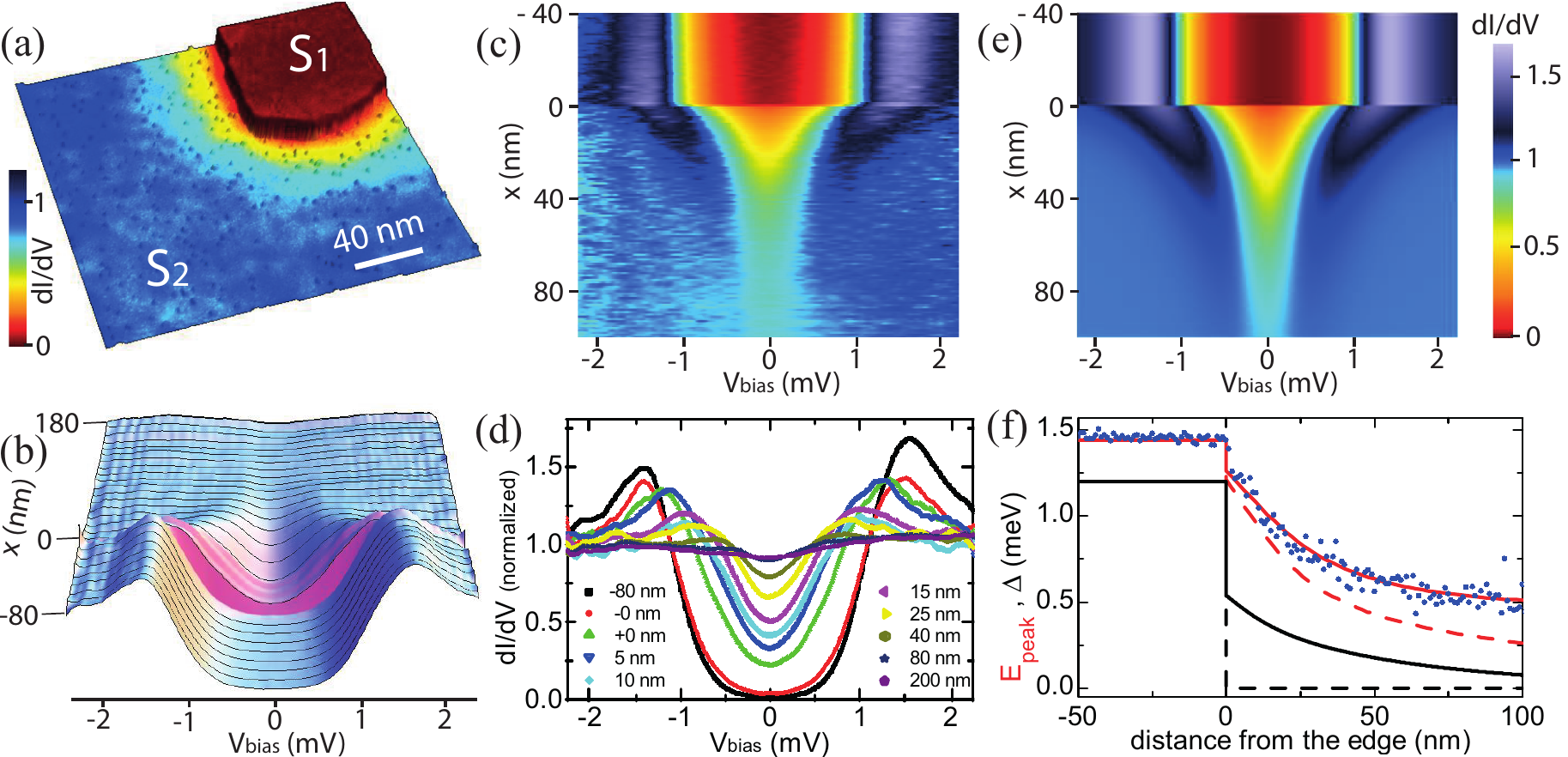}
\caption{The same as in Fig.~\ref{fig_exp_lowT}, but for $T=2.05$~K. At this temperature the striped-incommensurate
Pb monolayer $S_2$ is in its normal state. Notice that the order parameter determined self-consistently exibits
a finite value close the $S_1$-$S_2$ interface.}
\label{fig_exp_highT}
\end{figure*}

Let us now first discuss the results obtained at low temperature $T=0.3$~K $< T_{C1}, T_{C2}$ when both
electrodes are in the superconducting state. These results are summarized in Fig.~\ref{fig_exp_lowT}. In
particular, panel (a) displays the $S_1$-$S_2$ junction considered throughout this work, where $S_1$ is the
Pb island and $S_2$ corresponds to the SIC monolayer. The color-coded conductance map shown in this panel, 
at a bias voltage close to the $S_2$ gap edge, emphasizes that the proximity effect extends significantly 
far from the island edge. Let us stress that in Figs.~\ref{fig_exp_lowT}, \ref{fig_exp_highT}, and \ref{fig_IPE}, 
one single $dI/dV(V,x)$ spectrum corresponds to the average of all $dI/dV(V,x)$ spectra measured at the same 
distance $x$ from the island edge (the $S_1$-$S_2$ interface) in Fig.~\ref{topo_proxy}(a). As it can be seen 
from the extension of the superconducting correlations shown in Fig.~\ref{fig_exp_lowT}(a) and 
Fig.~\ref{fig_exp_highT}(a), this averaging procedure represents well the main behavior when going away from 
all island edges except in the vicinity of the sharp corner region, where the extension of the 
superconducting correlation is slightly reduced. Fig.~\ref{fig_exp_lowT}(b,c) shows 
the detailed evolution of the local $dI(V,x)/dV$ spectra as a function of the distance $x$ from the island edge 
measured in steps of 1~nm. Representative spectra are also shown in Fig.~\ref{fig_exp_lowT}(d). 
The main spectral features are the following [see labels in Fig.~\ref{fig_exp_lowT}(c)]: (A) Close to the 
interface, there is a proximity region where the spectra gradually evolve from one bulk behavior to the other 
over a distance of more than 100~nm. (B) A tiny yet important spectral feature is a small discontinuity in the 
height of the coherence peaks occurring at the $S_1$-$S_2$ interface on a subnanometer scale. (C,D) Far away 
from the island edge, the spectra go back to their bulk forms, $S_1$ ($S_2$) exhibiting a spatially constant 
superconducting gap $\Delta_1$ ($\Delta_2$). (E) Over $\sim 60$~nm from the $S_1$-$S_2$ interface, the spectra 
evolve in $S_1$ with slight changes toward the bulk $S_1$ spectrum, revealing the inverse proximity effect.

Before presenting how the tunneling conductance spectra are modified by increasing the temperature, it is 
convenient to introduce first the theoretical model that will help us to understand our observations. This is 
the goal of the next section. 

\section{Theoretical modeling: Usadel equations} \label{sec-III}

Most theoretical studies of the proximity effect in $S_1$-$S_2$ systems have focused on the analysis of their
critical temperature. These studies have made use either of the Ginzburg-Landau theory or of the linearized Gorkov's
equations \cite{Deutscher1969}, which are only valid close to critical temperature of the whole system \cite{note1}. 
Here, in order to describe the local spectra at arbitrary temperatures we used the Usadel approach \cite{Usadel1970}.
Usadel equations summarize the quasiclassical theory of superconductivity in the diffusive limit, where the mean
free path is smaller than the superconducting coherence length. In the monolayer $S_2$, this length is given
by $\xi_{2} = \sqrt{\hbar D_{2}/ \Delta_{2}}$. The quasiclassical theory describes all the equilibrium properties
in terms of a momentum averaged retarded Green's function $\hat G({\bf R},E)$, which depends on position ${\bf R}$
and energy $E$. This propagator is a $2\times 2$ matrix in electron-hole space
\begin{equation}
\label{eq-Gdef}
\hat G = \left( \begin{array}{cc} g & f \\
\tilde f & \tilde g \end{array} \right) .
\end{equation}
Neglecting inelastic and phase-breaking interactions, the propagator $\hat G({\bf R},E)$ satisfies the following
equation \cite{Usadel1970}
\begin{equation}
\label{usadel-eq}
\frac{\hbar D}{\pi} \nabla ( \hat G \nabla \hat G ) + [ E \hat \tau_3 +
\hat \Delta,  \hat G ] = 0 .
\end{equation}
\noindent
Here, $\hat \tau_3$ is the Pauli matrix in electron-hole space and
\begin{equation}
\label{eq-Deltadef}
\hat \Delta = \left( \begin{array}{cc} 0 & \Delta({\bf R}) \\
\Delta^{\ast}({\bf R}) & 0 \end{array} \right) ,
\end{equation}
where $\Delta({\bf R})$ is the space-dependent order parameter that needs to
be determined self-consistently via the following equation:
\begin{equation}
\label{eq-DeltaSC}
\Delta({\bf R}) = \lambda \int^{\epsilon_c}_{-\epsilon_c} \frac{dE}{2\pi} \mbox{Im}
 \left\{ f({\bf R},E) \right\} \tanh \left( \frac{\beta E}{2}
\right).
\end{equation}
Here, $\beta = 1/k_{\rm B}T$ is the inverse temperature, $\lambda$ is the coupling
constant, and $\epsilon_c$ is the cutoff energy. These two latter parameters
are eliminated in favor of the critical temperature of the monolayer (in the absence
of proximity effect) in the usual manner. Moreover, in our case, with no phase difference
between the superconducting reservoirs, the order parameter can be chosen real, as
we have done implicitly in Eq.~(\ref{eq-DeltaSC}).

To solve the Usadel equations in practice, we modeled our lateral $S_1$-$S_2$ system by a one-dimensional 
(1D) junction. Due to the big thickness difference between $S_1$ and $S_2$, we consider $S_1$ as an ideal 
reservoir in which the order parameter $\Delta_{1}$ remains constant and unperturbed up to the interface 
(the observed tiny deviations due to the inverse proximity effect in $S_1$ are discussed below). $S_2$
is approximated as a semi-infinite wire with a constant attractive pairing interaction $\lambda({\bf R}) =
\lambda_2$ that corresponds to a critical temperature $T_{C2}$. Following Ref.~\cite{Hammer2007}, we solved
the 1D Usadel equations using the Ricatti parametrization \cite{Eschrig2004} and described the junction
interface with Nazarov's boundary conditions, valid for arbitrary transparency \cite{Nazarov1999}. A key
parameter in these boundary conditions is an effective reflectivity coefficient, $r$, roughly defined as the
ratio between the resistances of the $S_1$-$S_2$ barrier and of the monolayer. Further technical details can be
found in Appendix A. Within our 1D model we compute the local DOS $\rho(x,E)$ as a function of the distance to 
the interface, $x$, as $\rho(x,E) = -\mbox{Im} \{ g(x,E) \}/\pi$, while the corresponding normalized tunneling 
spectrum is given by
\begin{equation}
\label{eq-dIdV}
\frac{dI}{dV}(x,V) = -\int^{\infty}_{-\infty} dE\, \rho(x,E)
\frac{\partial n_{\rm F}(E-eV)}{\partial E} .
\end{equation}
where $n_{\rm F}(E)$ is the Fermi function.

Let us now use this model to describe the low temperature results described in the previous section. 
For this purpose, we first fixed the bulk gaps in $S_1$ and $S_2$ by performing BCS fits of their 
local tunneling spectra acquired far away from $S_1$-$S_2$ interface. The best fits were obtained 
for $\Delta_{1} = 1.20$~meV and $\Delta_{2} = 0.23$~meV, with an effective electron temperature of 0.55~K 
slightly higher than the base temperature of our STM (see Appendix B). Second, we determined the value
of the effective reflectivity coefficient $r$ by adjusting the discontinuity in the spectra observed at
the interface. We obtained $r=0.02$, which implies a highly transparent yet non-perfect interface. Third,
we fixed the diffusion constant $D_{2}$ as to reproduce the spatial dependence of the energy of the spectral
maximum $E_{peak}(x > 0)$, see Fig.~\ref{fig_exp_lowT}(f). We obtained $D_{2} \approx 7.3$~cm$^2$/s, which
corresponds to a coherence length of $\xi_{2} \approx 45.7$~nm, in nice agreement with the $\xi_{2}$
value extracted from the analysis of the vortex core profile in the striped incomensurate phase 
\cite{Zhang2010,tobepublished}. Moreover, the value of $D_2$ suggests that the mean free path is rather small
(around 1 nm) and therefore, it is much smaller than $\xi_{2}$, which justifies the use of the Usadel
approach.

Fig.~\ref{fig_exp_lowT}(f) shows that the theory captures the decay length of $E_{peak}(x)$ along with the
observed jump at the island edge. In the same panel we also show the self-consistent order parameter $\Delta(x)$
in the Pb monolayer: It exhibits a jump at the island edge and decays gradually to the $S_2$ bulk value within
80-100~nm. If the order parameter is not accounted for self-consistently, the calculated spatial dependence
$E_{peak}(x)$ does not follow the experimental data, see dashed line in Fig.~\ref{fig_exp_lowT}(f). This
result emphasizes the need for a fully self-consistent calculation, which in any case is required based
on fundamental principles. In Fig.~\ref{fig_exp_lowT}(e) we show the $dI/dV(V,x)$ spectra obtained from the 
solution of the Usadel equations with the parameter values determined above. As one can see, the theory 
reproduces all the salient features of the experimental results of Fig.~\ref{fig_exp_lowT}(c,d).

\section{Giant proximity effect: Temperatures above $T_{C2}$} \label{sec-IV}

Let us now present and discuss the results for temperatures above the critical temperature of the monolayer.
At $T=2.05$~K, when $S_2$ is already in the normal state, the tunneling spectra change markedly, see
Fig.~\ref{fig_exp_highT}(a,b). Now, the spectra in the Pb monolayer close to the interface exhibit a smooth
induced gap which gradually disappears over a distance of $\sim 60$~nm away from the island edge. The overall
evolution of the spectra resembles that recently reported by us in a $S$-$N$ system, in which an amorphous
Pb wetting layer played the role of a 2D disordered normal metal \cite{Serrier-Garcia2013}. However, there
are two important differences with respect to the present case: (i) here the Altshuler-Aronov reduction of
the low-bias tunneling density of states, characteristic of electronic correlations, is absent and (ii) the
crystalline monolayer is superconducting at lower temperature while the disordered Pb wetting layer is not.

We now compare the proximity spectra with the results of our model using values of the parameters determined
above, \emph{i.e.}\ at 0.3~K. The effective temperature was taken equal to the bath temperature 2.05~K.
The computed tunneling spectra are presented in Fig.~\ref{fig_exp_highT}(e). Again, the theoretical results
reproduce qualitatively the experimental spectra of Fig.~\ref{fig_exp_highT}(c) with no adjustable parameters.
More importantly, as we show in Fig.~\ref{fig_exp_highT}(f), the Pb monolayer develops locally in the vicinity
of the interface a finite order parameter which survives over a distance of more than 100~nm. The impact of this
proximity-induced order parameter can be appreciated by comparing these results with a non-self-consistent
calculation where the order parameter is assumed to vanish at this temperature, which would correspond to
the situation where $S_2$ is a non-superconducting metal. Such a calculation shows that the induced gap extends 
over a much shorter distance inside the Pb monolayer as compared to the experimental dependence, see also 
Appendix C. This fact is illustrated in Fig.~\ref{fig_exp_highT}(f) where we show that the experimental data 
for $E_{peak}(x)$ are fitted much more satisfactorily by the self-consistent calculation. Thus, our results 
provide a clear evidence for the existence of the proximity-induced superconductivity in the interface region. 
This phenomenon was already discussed theoretically by de Gennes and co-workers in the early 1960's 
\cite{deGennes1964}, but to our knowledge no direct observation has ever been reported. It is worth mentioning
that the existence of this long-range or giant proximity effect has been suggested in the context of 
high-temperature superconductors based on the analysis of the supercurrent in trilayer Josephson junctions
\cite{Bozovic2004,Covaci2006}.

\begin{figure}[t]
\begin{center} \includegraphics*[width=0.95\columnwidth,clip]{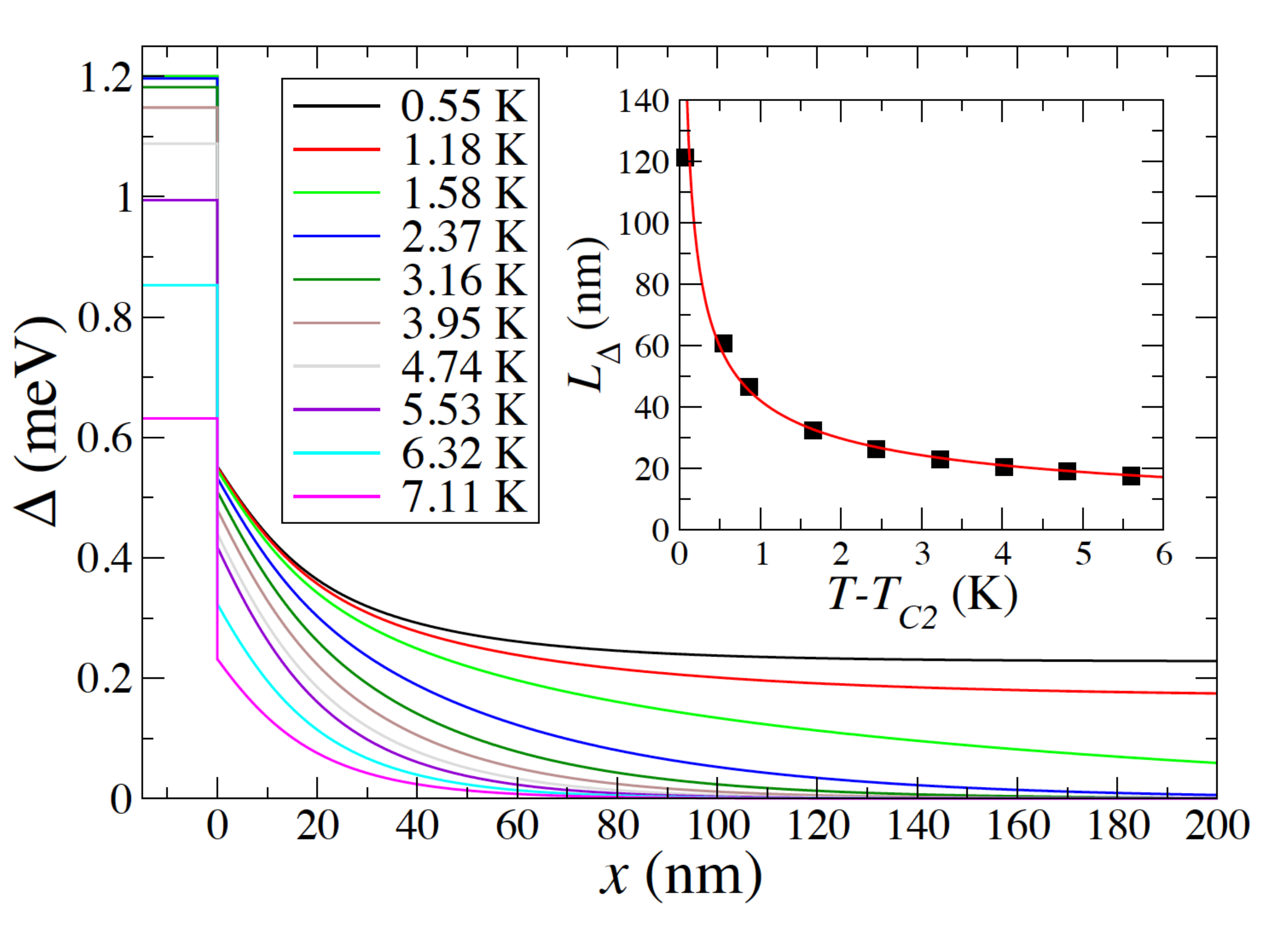} \end{center}
\caption{The main panel shows the computed order parameter in the Pb monolayer as a function of the distance
to the island edge for different temperatures. The critical temperature of the monolayer is within BCS theory
$T_{C2} = 1.5$1~K, while it is $T_{C1} = 7.89$~K for the island. The different parameters are those of
Fig.~\ref{fig_exp_highT}(f) and the temperature dependence of $\Delta_1$ has been taken into account. The
inset shows the temperature dependence of the decay length $L_{\Delta}$ of the induced order parameter
for temperatures above $T_{C2}$ (see text for definition). The symbols correspond to the values extracted
from the curves shown in the main panel, while the red solid line is a numerical fit of those results
to a function $\sim 1/\sqrt{T-T_{C2}}$.}
\label{op-vs-T}
\end{figure}

For completeness, we studied theoretically the proximity-induced interface superconductivity in a more
systematic manner. For this purpose, we computed the induced order parameter in $S_2$ for different temperatures,
see Fig.~\ref{op-vs-T}. For temperatures below $T_{C2}$, the order parameter tends to a finite value
far away from the island, while for higher temperatures it asymptotically vanishes. Notice that right above
$T_{C2}$, see light green curve in Fig.~\ref{op-vs-T}, the induced order parameter can extend inside the monolayer
several hundreds of nanometers. A detailed analysis of our numerical results in this temperature regime shows 
that away from the interface the induced order parameter decays exponentially as $\Delta(x) \propto 
\exp(-x/L_{\Delta})$, where $L_{\Delta}$ is a temperature-dependent decay length. As we illustrate in
the inset of Fig.~\ref{op-vs-T}, this decay length diverges when $T_{C2}$ is approached from above
approximately as $\sim 1/\sqrt{T-T_{C2}}$, which agrees with the prediction made with the help of the 
linearized Gorkov equations \cite{deGennes1964,Deutscher1969}. It is worth stressing that 
at distances $x < L_{\Delta}$, the induced order parameter does not follow this simple exponential decay. On the 
other hand, notice also that when the temperature approaches $T_{C1}$, both the island's order parameter and the 
induced one in the monolayer vanish altogether. Thus, our analysis shows that the proximity effect in $S_1$-$S_2$ 
junctions is much richer than in the $S$-$N$ case recently considered in Ref.~[\onlinecite{Kim2012}].

\section{Discussion and conclusions} \label{sec-V}

\begin{figure}[t]
\begin{center} \includegraphics[width=0.8\columnwidth,clip]{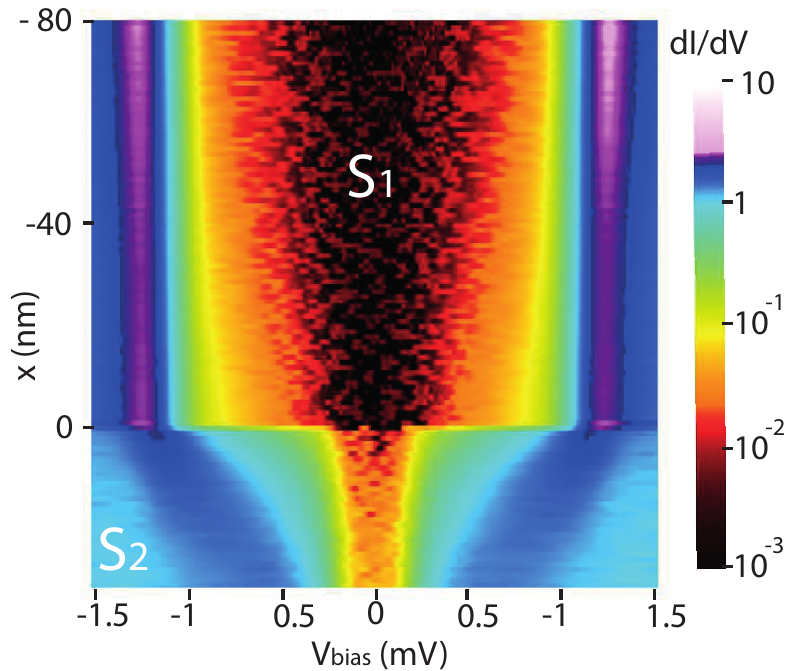} \end{center}
\caption{Inverse proximity effect in the $S_1$ island at 0.3~K. Color-coded experimental $dI/dV(V,x)$
spectra across the $S_1$-$S_2$ junction. One spectrum is plotted every nanometer. The four-decades log color scale
allows visualizing two important effects as one approaches the $S_1$-$S_2$ interface: (i) reduction of the peak 
energy and amplitude (pink narrow $dI/dV > 1$ band) and (ii) appearance of a tail of induced sub-gap 
states in the region $\Delta_2 \leq eV_{\rm bias} \leq \Delta_1$.}
\label{fig_IPE}
\end{figure}

It is worth stressing that we also observed the so-called inverse proximity effect in the Pb island, see 
Fig.~\ref{fig_IPE}, which we ignored for simplicity in our calculations. Indeed, although the Pb island is 
much closer to be an electron reservoir than the atomically thin Pb monolayer, the electron density ratio between 
$S_1$ and $S_2$ is not infinite, and the superconductivity in $S_1$ is also affected near the interface by a 
contact to a weaker superconductor $S_2$. Nevertheless the spectroscopic effects produced in $S_1$ are weak 
due to the large difference in electron densities of the two systems, and require a four-decades log scale 
to be clearly visible. Fig.~\ref{fig_IPE} allows us to see two important features of the inverse proximity 
effect: (i) Following the coherence peak heights represented by the violet-light pink bands (normalized 
conductance values above 1), one sees that the peak height slightly increases away from the $S_1$-$S_2$ 
interface while the energy of the peak maximum also slightly increases toward higher energy. (ii) Focusing 
now on the behavior of the low conductance values in the green-orange-brown region ($10^{-3}$-few $10^{-1}$), 
a tail of induced sub-gap states appears in the excitation spectrum of $S_1$ in the energy window $|E| \in 
[\Delta_{1}, \Delta_{2}]$. This tail of induced states enables to connect by the edges the large gap of 
$S_1$ to the small gap of $S_2$. In principle the inverse proximity effect can be described within a 
natural extension of our 1D model. However, as we discuss in Appendix D, such a description is not quite 
satisfactory and this limitation calls for an extension of our model that is presently in progress.

To summarize, we have presented an experimental study of the proximity effect between two different superconductors
$S_1$ and $S_2$. Our system consists of an \emph{in situ} fabricated lateral junction composed of a large-gap
Pb island $S_1$ ($\Delta_1=1.20$~meV) and a small-gap crystalline atomic Pb monolayer $S_2$ ($\Delta_2 =0.23$~meV).
Making use of a low-temperature STM/STS we have probed the local DOS of such a hybrid system with an unprecedented
energy and space resolution. The observed proximity-induced modification of the tunneling spectra in the Pb monolayer
$S_2$ was rationalized with the help of a 1D model based on the self-consistent solution of the Usadel equations. In 
particular, our results show the appearance of proximity-induced interface superconductivity in $S_2$ in the vicinity of
the $S_1$-$S_2$ interface for temperatures above $T_{C2}$, thus confirming the theoretical prediction by de Gennes and
co-workers \cite{deGennes1964,Deutscher1969}. Our work not only elucidates this old-standing problem in the context 
of proximity effect, but it also paves the way for studying new aspects of this quantum phenomenon such as the Meissner 
effect and vortex phases in proximity-induced superconductors. Finally, a weak inverse proximity effect is 
also revealed in the $S_1$ Pb island, characterized by both a slight reduction of the coherence peak height and 
energy and by the appearance of a tail of induced sub-gap states. Further generalization of our 1D model is needed 
to account for the inverse proximity effect.

\begin{acknowledgments}
We thank Hugues Pothier, Cristian Urbina, and Guy Deutscher for useful discussions. This work was supported
by an ``Emergence" grant from the University Pierre et Marie Curie (UPMC) and by a CFM foundation PhD grant
(G.M.). V.S.\ acknowledges financial support from the Ministry of Education and Science of the Russian Federation
and J.C.C.\ from the Spanish MICINN (Contract No.\ FIS2011-28851-C02-01).
\end{acknowledgments}

\appendix

\section{Theoretical description of the proximity effect: Usadel equations}

Our approach to describe the proximity effect is based on the Usadel equations \cite{Usadel1970}. The goal 
of this appendix is to provide some additional technical details about how they were solved in practice. 
In this discussion, we shall follow closely Ref.~[\onlinecite{Hammer2007}].

To describe the proximity effect in the crystalline monolayer, we modeled the Pb islands as ideal superconducting
reservoirs and the monolayer as an infinite normal wire. To implement this model in practice, we considered a
1D $S_1$-$S_2^{\prime}$-$S_2$ junction, where $S_1$ and $S_2$ are BCS superconducting reservoirs
with constant gaps $\Delta_{\rm island} = \Delta_1$ and $\Delta_{\rm monolayer} = \Delta_2$, respectively, and
$S_2^{\prime}$ is a superconducting wire of the same material as the monolayer reservoir. The length of
the central superconducting wire, $L$, was chosen sufficiently large as to avoid the influence of the
presence of the $S_2$ reservoir in the density of states close to the interface with the $S_1$ reservoir.
On the other hand, we neglected the inverse proximity effect in the $S_1$ reservoir (island) and we assumed that 
the $S_2^{\prime}$-$S_2$ interface was perfectly transparent. However, we allowed the $S_1$-$S_2^{\prime}$ to
be non-ideal, as we explain in more detail below.

As explained in section \ref{sec-III}, our technical task is to solve the Usadel equations 
[Eq.~(\ref{usadel-eq})] together with the corresponding equation for the order parameter [Eq.~(\ref{eq-DeltaSC})]. 
Equation (\ref{usadel-eq}) must also be supplemented by the normalization condition $\hat G^2 = -\pi^2 \hat 1$. 
In order to solve numerically the Usadel equation it is convenient to use the so-called Riccati parameterization 
\cite{Eschrig2000}, which accounts automatically for the normalization condition. In this case, the retarded Green's
functions are parameterized in terms of two coherent functions $\gamma({\bf R},E)$ and $\tilde \gamma({\bf R},E)$ 
as follows
\begin{eqnarray}
\label{riccati}
\hat G & = & - \frac{i \pi}{1 + \gamma \tilde \gamma} \left( \begin{array}{cc}
1 - \gamma \tilde \gamma & 2 \gamma \\
2 \tilde \gamma & \gamma \tilde \gamma -1 \end{array} \right) .
\end{eqnarray}

Using their definition in Eq.~(\ref{riccati}) and the Usadel equation (\ref{usadel-eq}), one can obtain the 
following transport equations for these functions in the wire region \cite{Eschrig2004}
\begin{eqnarray}
\label{g-eq}
\partial^2_{\tilde x} \gamma + \frac{\tilde f}{i\pi} (\partial_{\tilde x} \gamma)^2 +
2i \left( \frac{E}{E_{\rm T}} \right) \gamma & = & - i \frac{\Delta}{E_{\rm T}}(1+ \gamma^2), \\
\partial^2_{\tilde x} \tilde \gamma + \frac{f}{i\pi} (\partial_{\tilde x} \tilde \gamma)^2 +
2i \left( \frac{E}{E_{\rm T}} \right) \tilde \gamma & = & i \frac{\Delta}{E_{\rm T}}(1+ \tilde \gamma^2) ,
\label{gtilde-eq}
\end{eqnarray}
where we have used that the order parameter is real. Here, $\tilde x$ is the dimensionless coordinate that describes 
the position along the $S_2^{\prime}$ wire and ranges from 0 ($S_1$ lead) to 1 ($S_2$ lead) and $E_{\rm T} = 
\hbar D_2/L^2$ is the Thouless energy of the wire. The expressions for $\tilde f$ and $f$ are obtained by 
comparing Eq.~(\ref{eq-Gdef}) with Eq.~(\ref{riccati}). Notice that Eqs.~(\ref{g-eq}) and (\ref{gtilde-eq}) 
couple the functions with and without tilde. However, for the system under study one can show that the symmetry 
$\tilde \gamma({\bf R},E) = - \gamma({\bf R},E)$ holds and therefore, only Eq.~(\ref{g-eq}) needs to be solved.

Now, we have to provide the boundary conditions for Eq.~(\ref{g-eq}). Since we want to describe a semi-infinite 
region of the striped incommensurate phase, we assume that the $S_2^{\prime}$-$S_2$ interface is perfectly 
transparent. In this case, the boundary condition in this interface is simply given by the continuity of the 
coherent function:
\begin{equation}
\label{eq-gamma-res}
\gamma(\tilde x=1,E) = \gamma_2(E) = -\frac{\Delta_2} {E^R + i \sqrt{\Delta^2_2 -(E^R)^2}},
\end{equation}
with $E^R = E+i0^+$.

We allow the interface between the island and the monolayer wire to be non-ideal and to describe it, we used 
the boundary conditions derived in Refs.~[\onlinecite{Nazarov1999,Kopu2004}]. These conditions for an spin-conserving
interface are expressed in terms of the Green's functions as follows
\begin{equation}
\hat G_{\beta} \partial_{\tilde x} \hat G_{\beta} = \left( \frac{G_0}
{G_{\rm N}} \right) \sum_i \frac{2\pi^2 \tau_i \left[ \hat G_{\beta},
\hat G_{\alpha} \right]} {4\pi^2 - \tau_i \left(
\left\{ \hat G_{\beta}, \hat G_{\alpha} \right\} + 2\pi^2 \right)} .
\end{equation}
Here, $\hat G_{\beta(\alpha)}$ refers to the Green's function on the monolayer (island) side of the interface, 
$G_0=2e^2/h$ is the quantum of conductance, and $\tau_i$ are the different transmission coefficients characterizing 
the interface.  The parameter $G_{\rm N}$ is equal to $\sigma_2 S/L$, where $\sigma_2$ is the normal-state conductivity of 
the monolayer and $S$ is the cross section of the barrier. Thus, $G_{\rm N}$ can be viewed as the normal-state conductance 
of a wire with cross section $S$ and length $L$. In general, one would need the whole set $\{ \tau_i \}$, but since 
one does not have access to this information we adopt here a practical point of view. We assume that all the $M$ 
interface open channels have the same transmission $\tau$ and define $G_{\rm B} = G_0 M \tau$ as the conductance 
of the barrier. Thus, the $S_1$-$S_2^{\prime}$ interface is characterized by two quantities, namely the barrier 
conductance $G_{\rm B}$ and the transmission $\tau$, and our starting point for the boundary conditions is
\begin{equation}
\label{non-ideal}
r \hat G_{\beta} \partial_{\tilde x} \hat G_{\beta} = \frac{2\pi^2 \left[ \hat G_{\beta},
\hat G_{\alpha} \right]} {4\pi^2 - \tau \left( \left\{ \hat G_{\beta},
\hat G_{\alpha} \right\} + 2\pi^2 \right)} ,
\end{equation}
where we have defined the ratio $r=G_{\rm N}/G_{\rm B}$. In this language, an ideal interface is characterized 
by $r=0$ and a tunnel contact is described by $\tau \ll 1$.

The next step is to express these boundary conditions directly in terms of the coherent functions. Substituting 
the definitions of Eq.~(\ref{riccati}) into Eq.~(\ref{non-ideal}) and after straightforward algebra, one obtains 
the following boundary conditions for the coherent function $\gamma$ at $\tilde x=0$:
\begin{eqnarray}
\label{g-boundary}
-r \frac{\partial_{\tilde x} \gamma_{\beta} - (\gamma_{\beta})^2 \partial_{\tilde x}
\gamma_{\beta}} {(1 - (\gamma_{\beta})^2)^2} & = & \nonumber \\
\frac{(1 + (\gamma_{\beta})^2) \gamma_{\alpha} - (1 + (\gamma_{\alpha})^2 )
\gamma_{\beta}} {(1 - (\gamma_{\beta})^2) (1 - (\gamma_{\alpha})^2) +
\tau (\gamma_{\alpha} - \gamma_{\beta})^2} . &&
\end{eqnarray}
This equation establishes a relation between the coherent function and its derivative evaluated on the side 
of the interface inside the $S_2^{\prime}$ wire ($\beta$) and the corresponding function evaluated on the side 
of the interface inside the $S_1$ reservoir ($\alpha$), which is given by Eq.~(\ref{eq-gamma-res}) replacing 
$\Delta_2$ by $\Delta_1$.

In summary, our main task was to solve Eq.~(\ref{g-eq}) together with Eq.~(\ref{eq-Deltadef}) in a self-consistent 
manner. The non-linear second order differential equation of Eq.~(\ref{g-eq}), together with its boundary conditions 
of Eqs.~(\ref{eq-gamma-res}) and (\ref{g-boundary}), is a typical two point boundary value problem. We solved it
numerically using the so-called relaxation method as described in Ref.~[\onlinecite{relaxation}]. On the other 
hand, the self-consistent loop was done using a simple iterative algorithm until convergence was achieved. Finally, 
the numerical solution for the coherent function was used to construct the retarded Green's function from 
Eq.~(\ref{riccati}) and to compute the local density of states in the striped incommensurate phase region and
the corresponding local tunneling conductance, see Eq.~(\ref{eq-dIdV}).

\begin{figure}[b]
\begin{center} \includegraphics[width=\columnwidth,clip]{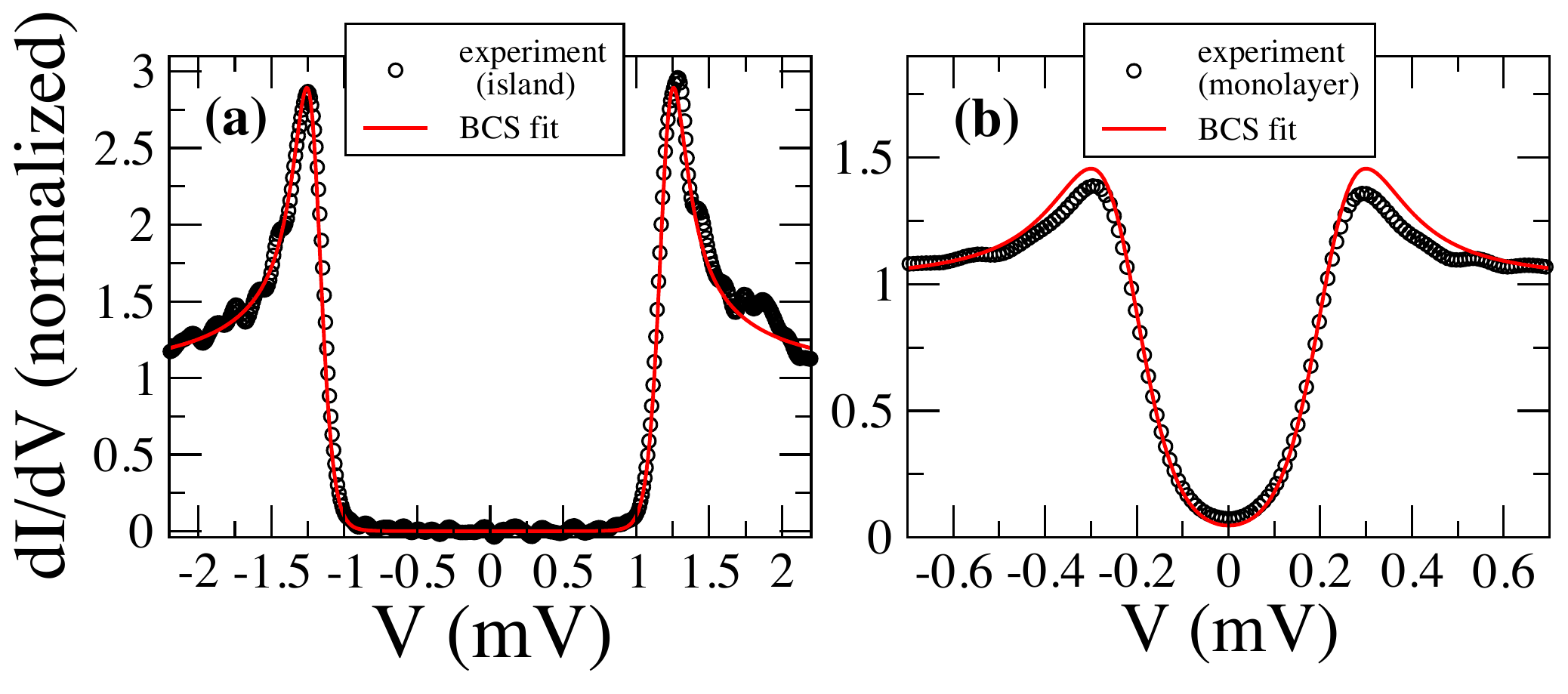} \end{center}
\caption{Determination of the zero-temperature bulk gaps. (a) Normalized tunneling
spectra as a function of the bias voltage. The symbols correspond to the experimental
spectrum measured at 0.3~K deep inside the Pb island ($x=-80$~nm). The solid line
corresponds to the best fit obtained with the BCS theory for $\Delta_1 = 1.2$~meV and
$T=0.55$~K. (b) The same as in the upper panel but for a spectrum measured in the
Pb monolayer very far away from the island edge ($x=+200$~nm). The BCS tunneling
spectrum was obtained using $\Delta_2 = 0.23$~meV and $T=0.55$~K.}
\label{BCS-fits}
\end{figure}

\section{Bulk gaps: BCS fits}

A first necessary step in the explanation of our experimental results is the determination
of the zero-temperature bulk gaps of the island ($\Delta_1$) and the Pb monolayer ($\Delta_2$).
For this purpose, we fitted the tunneling spectra measured at 0.3~K deep inside both
superconductors with the bulk BCS theory, \emph{i.e.}\ using the bulk BCS DOS in Eq.~(\ref{eq-dIdV}).
To do the fits we used both the zero-temperature gap and the temperature as adjustable parameters.
The results of these fits are shown in Fig.~\ref{BCS-fits}. As one can see, the BCS theory
reproduces satisfactorily the bulk spectra of both superconductors with an effective temperature
of 0.55~K, which is slightly higher than the bath temperature of our experiments. The values
obtained for the zero-temperature gaps are: $\Delta_1 = 1.2$~meV and $\Delta_2 = 0.23$~meV. It
is worth stressing that no artificial broadening was introduced in the expression of the BCS DOS
to perform the fits.

\section{Theoretical results: Representative examples and the role of the self-consistency}

We are not aware of any theoretical work on the proximity effect between two diffusive superconductors
for arbitrary temperatures. For this reason, it may be of interest for future reference to provide 
here a more in depth discussion of the results of the model described above. 

\begin{figure}[t]
\includegraphics*[width=\columnwidth,clip]{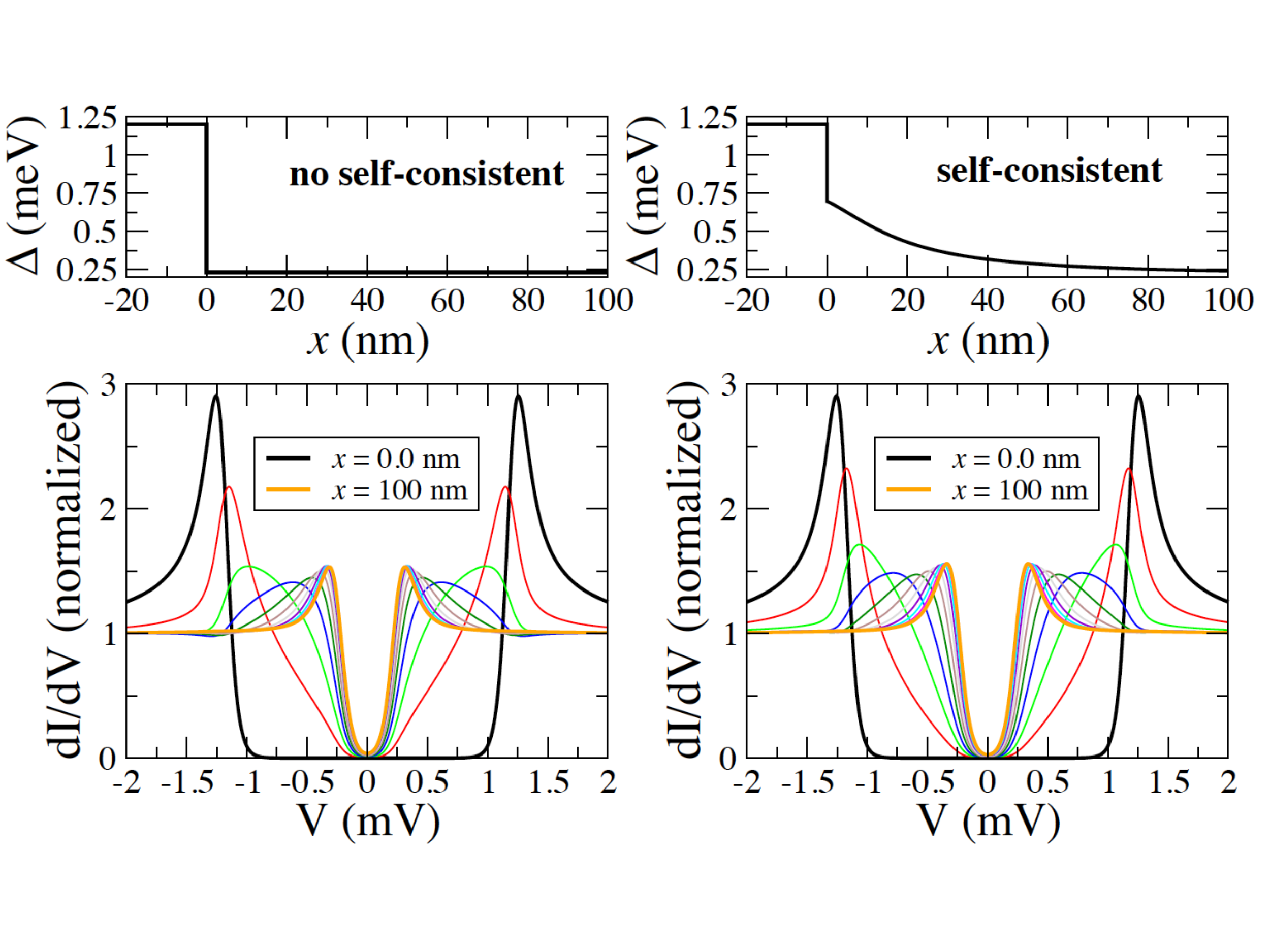}
\caption{The lower panels show the computed normalized $dI/dV$ as a function of the bias voltage
for $T=0.55$ K and $r=0.0$ (perfect transparency). The different curves correspond to
different positions along the monolayer, $x$. The curves were computed in steps of
10 nm ranging from $x=0$ (island edge) to $x=100$ nm. The lower left panel corresponds to
a calculation were the order parameter in the monolayer was assumed to be constant
and equal to its bulk value. The lower right panel corresponds to the case in which the
order parameter has been calculated in self-consistent manner. The upper panels show
the corresponding order parameter profiles.}
\label{cond-T=0.55K-r=0}
\end{figure}

In what follows, we fix the values of the zero-temperature gap of both superconductors
to the values that describe best the experimental results, \emph{i.e.}\ $\Delta_{\rm island} =
\Delta_1 = 1.2$~meV and $\Delta_{\rm monolayer} = \Delta_2 = 0.23$~meV, corresponding to critical
temperatures within the BCS theory equal to $T_{C1}=7.89$~K and $T_{C2}=1.51$~K. On the other
hand, we also keep fixed the value of the diffusion constant in the crystalline monolayer to
$D_2 = 7.3$~cm$^2$/s, which corresponds to a coherent length of $\xi_2 = 45.7$~nm. Let us start
our analysis by considering the case in which the interface between the island and the monolayer
is perfectly transparent ($r=0$). Assuming a temperature of 0.55~K, which corresponds to the lowest
effective temperature of the experiments, we have computed the local tunneling spectra in the monolayer
as a function of the bias voltage for different positions along the monolayer, $x$, and the
results can be seen in Fig.~\ref{cond-T=0.55K-r=0}. Here, we show both the results obtained assuming
a constant order parameter in the monolayer (see lower left panel), \emph{i.e.}\ ignoring the
self-consistency, and the full self-consistent results (see lower right panel). We also
show in the upper panels the corresponding profiles of the order parameter. The different
curves in the lower panels correspond to different distances to the island edge ranging from
0 to 100~nm. The curves have been calculated in steps of 10~nm. As one can see, the spectra
evolve gradually from the BCS-like spectrum at the island edge ($x=0$~nm) for $\Delta =
1.2$~meV to the BCS-like spectrum deep inside the monolayer ($x=100$~nm) for $\Delta =
0.23$~meV. In the vicinity of the island, the maximum of the spectra occurs in the range between
the two bulk gaps (slightly shifted by the finite temperature) and it approaches $\Delta_2$
at $\sim 2 \xi_2$. Notice, however, that the spectra exhibit a space-independent gap equal
to $\Delta_2$. This agrees with the results obtained by de Gennes using a variational method
\cite{deGennes1964}. Notice also that the self-consistency increases the magnitude of the bias
at which the spectra reach their maximum close to the interface, which is a consequence
of the larger order parameter in that region.

\begin{figure}[t]
\includegraphics*[width=\columnwidth,clip]{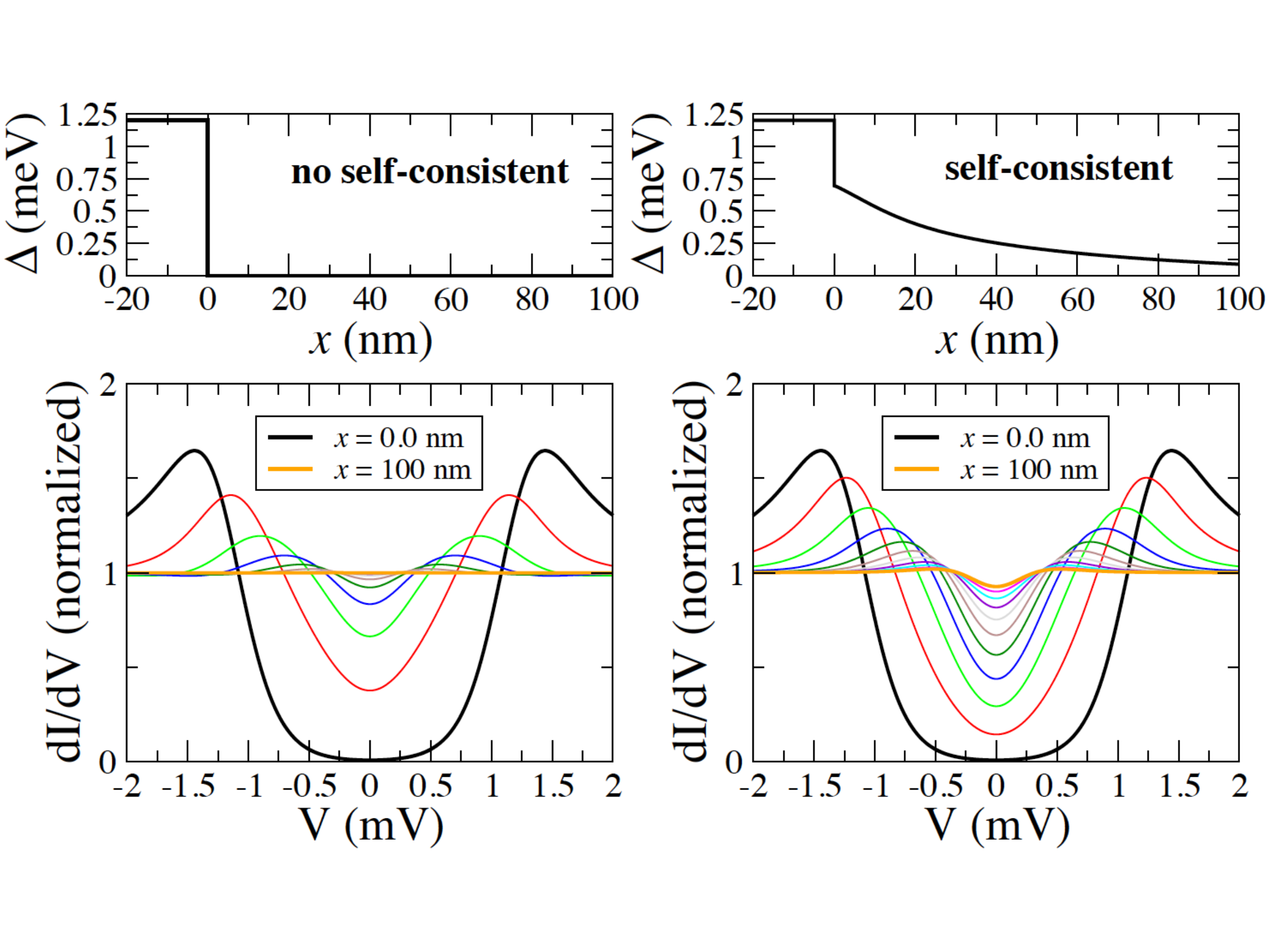}
\caption{The same as in Fig.~\ref{cond-T=0.55K-r=0}, but for $T=2.05$ K. At this temperature the
monolayer is in its normal state. Notice that the order parameter determined self-consistently
exhibits a finite value in the vicinity of the island.}
\label{cond-T=2.05K-r=0}
\end{figure}

Let us now consider a temperature of 2.05~K, which is above the critical temperature of the striped 
incommensurate monolayer. The corresponding results are shown in Fig.~\ref{cond-T=2.05K-r=0}. In this 
case, one may naively expect the system to behave like an ordinary $S$-$N$ system. However, this is 
not the case. As we show in the upper right panel of Fig.~\ref{cond-T=2.05K-r=0}, the monolayer 
develops a finite order parameter close to the interface due to the proximity effect. This fact has 
a strong impact in the spectra. In particular, the induced gap that naturally appears in a $S$-$N$ 
junction, see the non-self-consistent calculation in the lower left panel, now extends up to a much 
larger distance inside the monolayer giving rise to a long-range proximity effect. As we have also
discussed in section \ref{sec-IV}, see Fig.~\ref{op-vs-T}, this phenomenon persists up to the critical
temperature of the island, which in this case coincides with the critical temperature of the hybrid
structure.

\begin{figure}[t]
\includegraphics*[width=\columnwidth,clip]{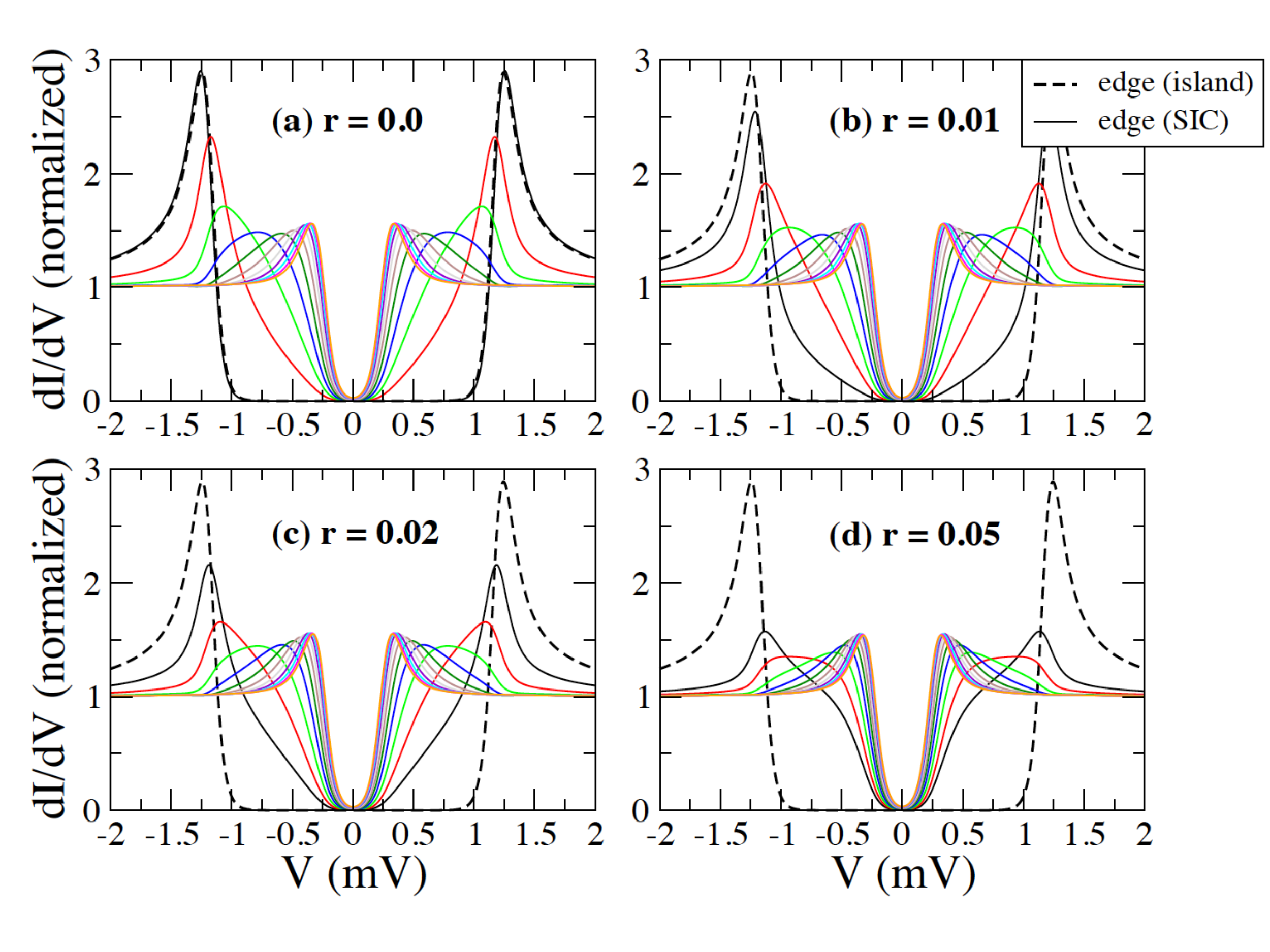}
\caption{Normalized tunneling conductance at $T=0.55$ K as a function of the bias voltage for different
values of the interface parameter $r$, as indicated in the panels. The different curves correspond
to different positions along the monolayer, $x$. The curves were computed in steps of 10 nm ranging from
$x=0$ (island edge) to $x=100$ nm. The black dashed lines in the different panels correspond to the
spectra inside the island, while the black solid lines are the results obtained in the monolayer side
of the island edge. Notice the discontinuity in the spectra at the island edge when $r \ne 0$.}
\label{cond-T=0.55K-vs-r}
\end{figure}
\begin{figure}[b]
\includegraphics*[width=\columnwidth,clip]{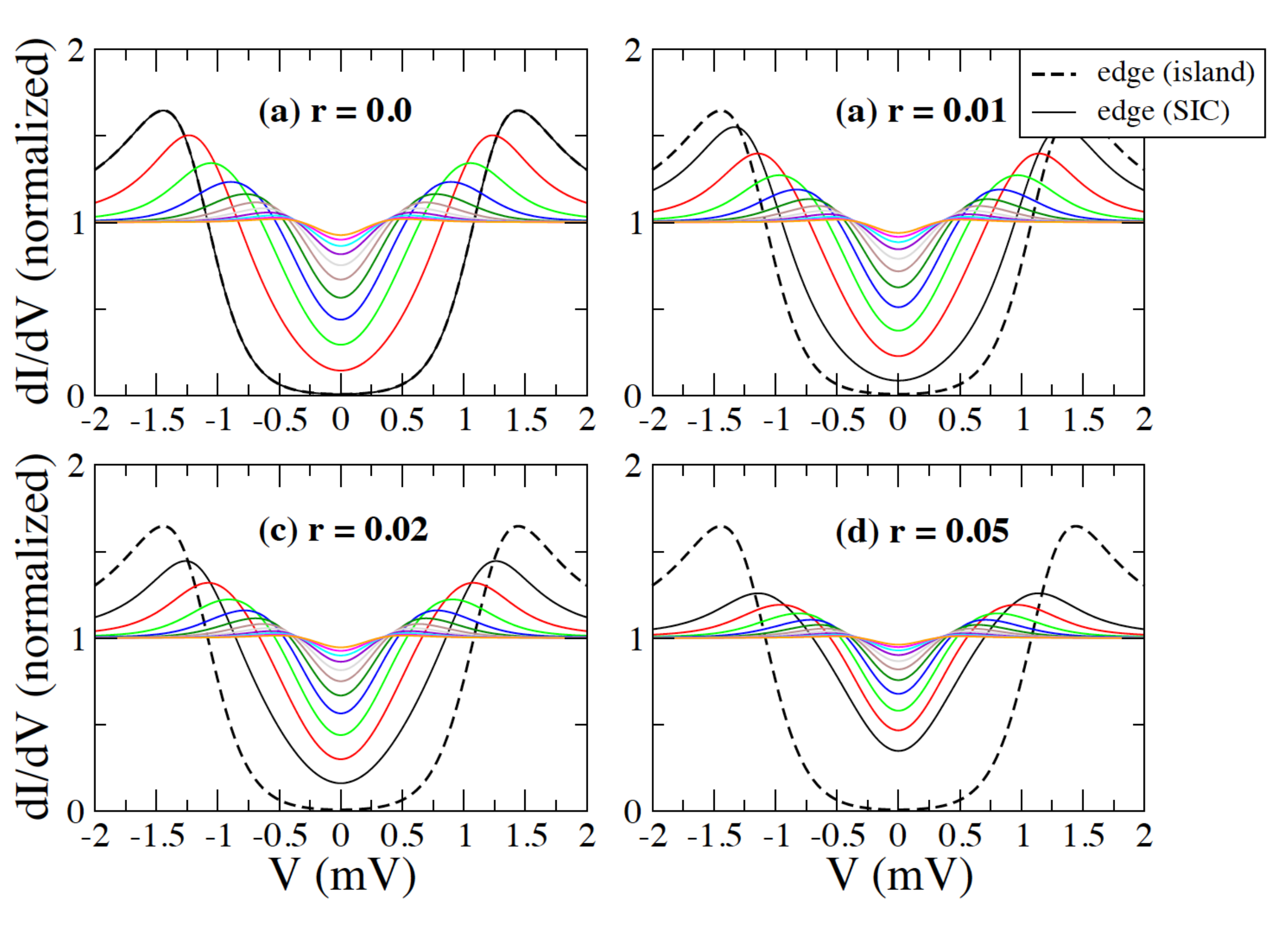}
\caption{The same as in Fig.~\ref{cond-T=0.55K-vs-r}, but for $T=2.05$~K.}
\label{cond-T=2.05K-vs-r}
\end{figure}

Irrespective of the temperature, in the previous results the spectra evolve continuously 
from the island to the crystalline monolayer. This is a consequence of the assumed
perfect transparency. However, the experimental results discussed in the main text show that
there is a jump in the spectra when crossing the island edge. This suggests that the
interface, although highly transparent, it is not perfect. This discontinuity can be
accounted for using the boundary conditions described above. For simplicity, we assume
that the transmission coefficient $\tau$ is equal to 1 and we attribute the non-ideality
of the interface to a non-negligible resistance of the barrier ($r \ne 0$). To illustrate
the role of a finite value of the parameter $r$, we computed the tunneling spectra for
different values of $r$ for 0.55 and 2.05~K. The results are displayed in
Figs.~\ref{cond-T=0.55K-vs-r} and \ref{cond-T=2.05K-vs-r} and, as one can see, a
finite $r$ produces two main effects. First, there appears a jump at the island edge which
increases in magnitude as $r$ increases. Indeed, we have used the magnitude of this jump
to adjust the value of $r$. Second, the peaks height in the spectra in the vicinity
of the island decreases as the value of $r$ increases. If we kept increasing the value of $r$,
the spectra in the monolayer would tend to be constant and we would recover the result
for a bulk superconductor with its corresponding bulk gap (0.23~meV at 0.55~K and 0~meV at
2.05~K).

\begin{figure}[t]
\begin{center} \includegraphics*[width=0.7\columnwidth,clip]{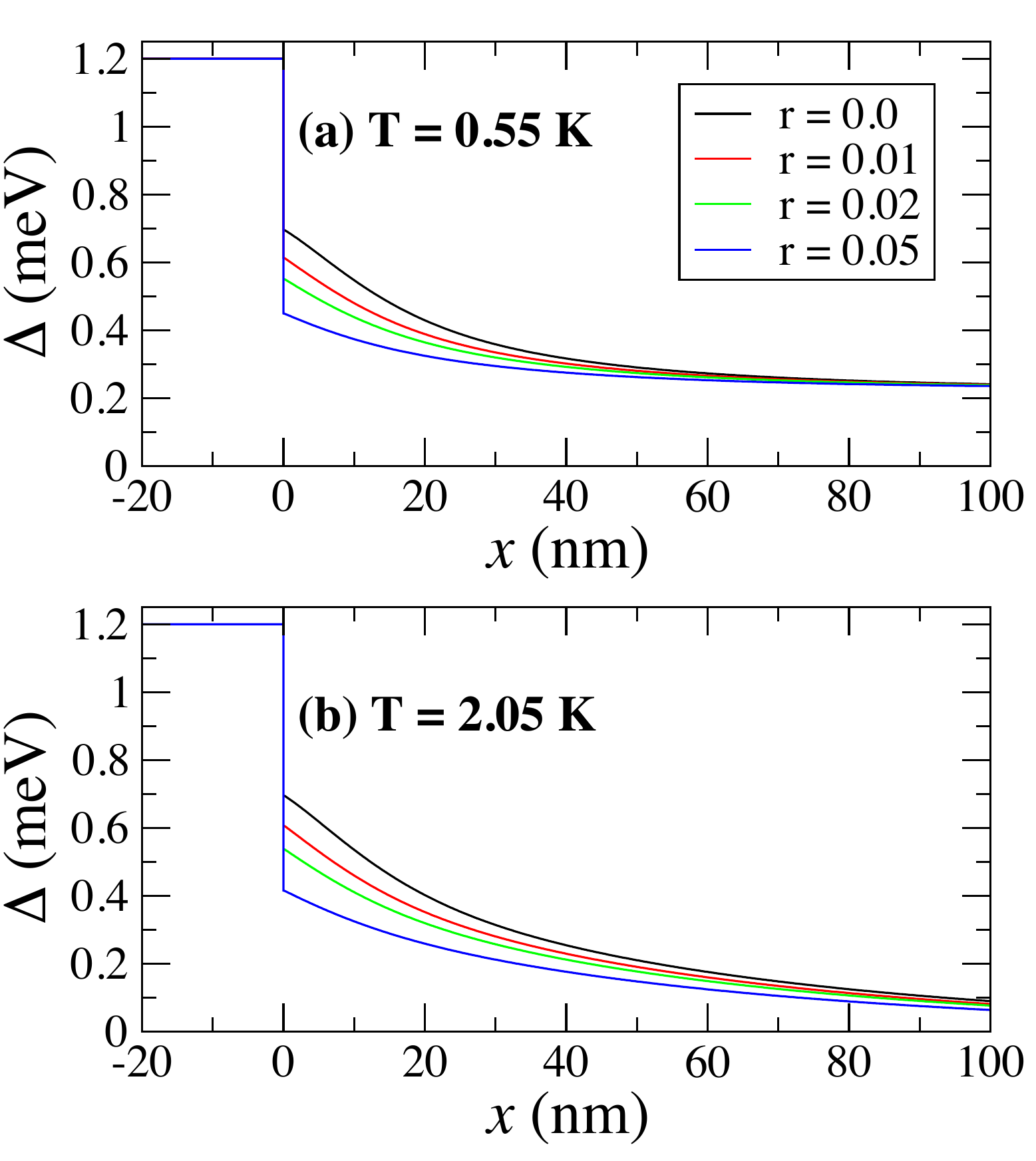} \end{center}
\caption{Self-consistent order parameter as a function of the position for different values of
$r$ and for temperatures (a) $T=0.55$~K and (b) $T=2.05$~K. These profiles correspond to the
cases shown in Fig.~\ref{cond-T=0.55K-vs-r} and \ref{cond-T=2.05K-vs-r}.}
\label{op-vs-r}
\end{figure}

The finite interface resistance ($r \ne 0$) has also an obvious influence in the profiles of
the order parameter. In Fig.~\ref{op-vs-r} we show the order parameter profiles corresponding
to the different cases considered in Figs.~\ref{cond-T=0.55K-vs-r} and \ref{cond-T=2.05K-vs-r}.
As one can see, a finite value of $r$ reduces the amplitude of the order parameter in the vicinity
of the island, reducing so the proximity effect. Again, a very high value of $r$ would simply
kill the proximity effect in the monolayer.

\begin{figure}[t]
\begin{center} \includegraphics*[width=0.8\columnwidth,clip]{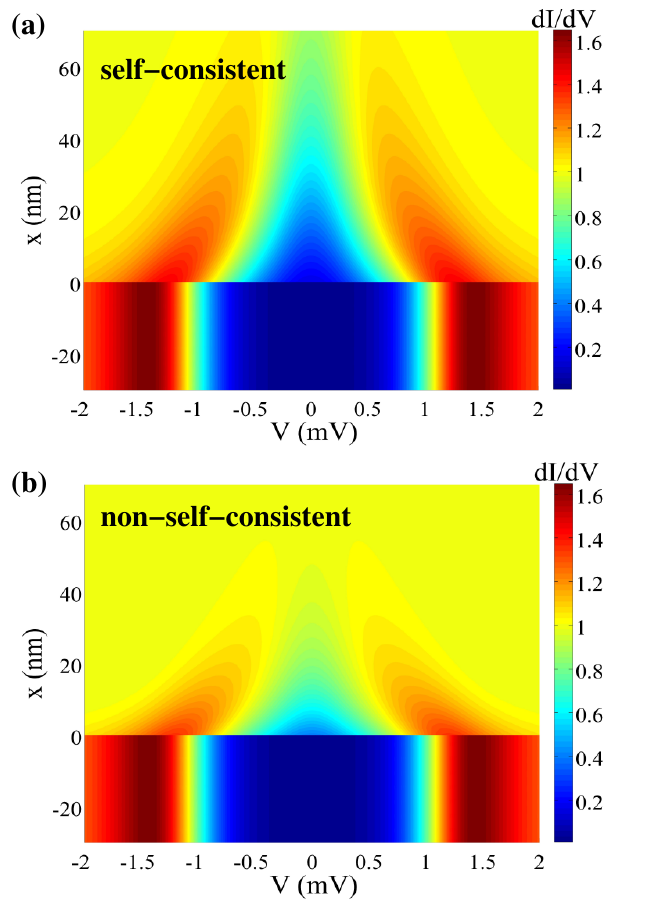} \end{center}
\caption{Computed normalized tunneling spectra as a function of the bias and the position measured
with respect to the island edge. (a) Results obtained in a self-consistent calculation using the
following parameter values: $T=2.05$~K, $\Delta_1 = 1.2$~meV, $\Delta_2 = 0.23$~meV, $r = 0.02$,
and $D_2 = 7.3$~cm$^2$/s. (b) The same as in panel (a), but for a non-self-consistent calculation
where the order parameter was assumed to vanish in the monolayer.}
\label{comp-sc-nosc}
\end{figure}

A crucial point in our discussion is the marked difference between the self-consistent and
non-self-consistent calculations, which is particularly clear for temperatures above the critical
temperature of the monolayer. This was already illustrated in Fig.~\ref{fig_exp_highT}(f) where we show
that the experimental evolution of $E_{peak}(x)$ at 2.05~K is much more satisfactorily described by the
self-consistent result. For completeness, we show in Fig.~\ref{comp-sc-nosc} a comparison between
the self-consistent and non-self-consistent calculations of the evolution of the tunneling spectra
at 2.05~K in a 2D plot for the same parameters as in Fig.~\ref{fig_exp_highT}(e,f). Notice that the color
scale is different from the one used in the main text. As one can see, in the non-self-consistent
case the proximity effect extends a much smaller distance inside the monolayer. This is a natural
consequence of the non-vanishing order parameter in the vicinity of the island that is found in the
self-consistent calculation. This induced order parameter is reflected in a long-range proximity
effect, as compared to the standard $S$-$N$ case where $N$ is a non-superconducting metal.

\begin{figure}[t]
\begin{center} \includegraphics*[width=\columnwidth,clip]{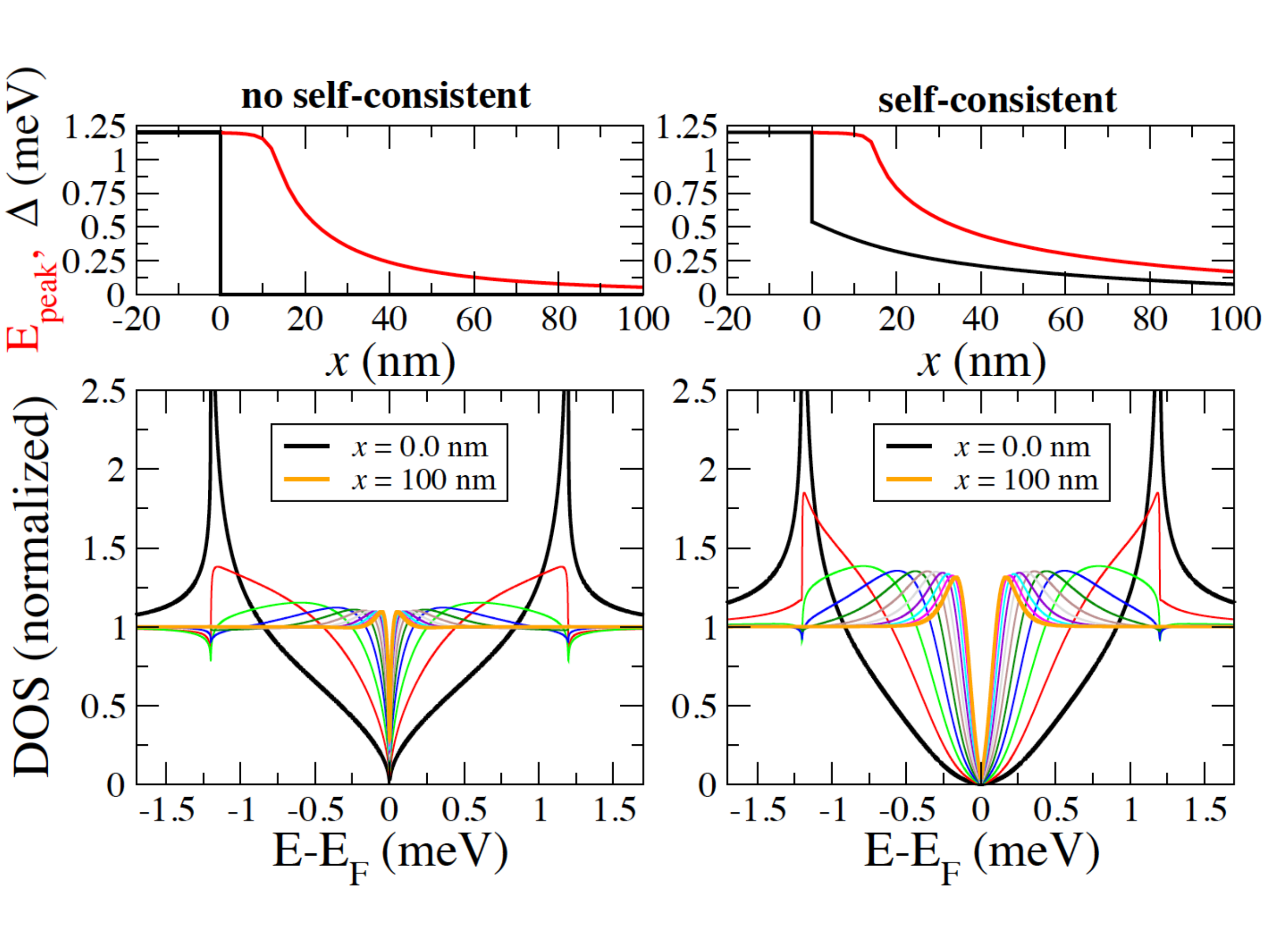} \end{center}
\caption{The lower panels show the computed normalized density of states as a function of energy
(measured with respect to the Fermi energy, $E_{\rm F}$). The parameters of the model were those
of Fig.~\ref{comp-sc-nosc} and the different curves correspond to different positions along the
monolayer, $x$. The curves were computed in steps of 10 nm ranging from $x=0$ (monolayer side of
the island edge) to $x=100$ nm. The lower left panel corresponds to a calculation were the order
parameter in the monolayer was assumed to be zero everywhere, while the lower right panel
corresponds to the case in which the order parameter was calculated in self-consistent manner.
The upper panels show the corresponding order parameter profiles (black lines) and the spatial
evolution of the coherent peaks (red lines).}
\label{DOS-thy}
\end{figure}

This long-range proximity effect can be better appreciated by looking directly at the local DOS, 
\emph{i.e.}\ getting rid of the thermal broadening of the tunneling conductance. For this reason,
in the lower panels of Fig.~\ref{DOS-thy} we show a comparison of the spatial dependence of the 
local DOS for the cases shown in Fig.~\ref{comp-sc-nosc}. We also include in the upper panels the
corresponding profiles of the order parameter, as well as the evolution of the coherent peaks,
defined as the position in energy of the maxima of the local DOS. As one can see in the lower
panels, in both cases the main feature in the local DOS is the appearance of an induced gap which progressively
disappears as we move away from the interface; an induced gap that is obviously more pronounced than 
in the conductance spectra. In the non-self-consistent case, which corresponds to a standard $S$-$N$
junction (with $N$ being a non-superconducting metal), the decay of the induced gap away from the 
interface can be fitted by a function of the form $\sim 1/(1 + x/L_{\xi})^2$, with $L_{\xi} = 14.3$ nm. 
This means that this induced gap roughly scales as a local Thouless energy $\hbar D_2/ x^2$, as it 
is well-known in standard diffusive $S$-$N$ junctions \cite{Belzig1996}. In the self-consistent case, 
the decay of the induced gap can also be accurately described with the same type of function, but this 
time the existence of a finite induced order parameter in the monolayer is manifested in a larger decay 
length, which we numerically found to be $L_{\xi} = 55.7$ nm for this example.

\section{Inverse proximity effect}

So far, we have assumed in our theoretical analysis that the island is a perfect reservoir with a
constant order parameter. However, as discussed in section \ref{sec-V}, our experiments show a small 
inverse proximity effect in the form of a reduction of the coherence peak amplitude and energy, as well 
as the appearence of subgap states in the excitation spectrum of $S_1$ in the energy window $|E| \in 
[\Delta_{1}, \Delta_{2}]$. It is straightforward to extend the model discussed above to try to describe 
this inverse proximity effect. This simply requires to solve the Usadel equation also inside the island 
taking into account that the pairing interaction constant, $\lambda$, takes different values in both 
electrodes, according to the corresponding critical temperatures. Moreover, we must consider that the
diffusion constant can be different in both superconductors.

\begin{figure*}[t]
\begin{center} \includegraphics*[width=\textwidth,clip]{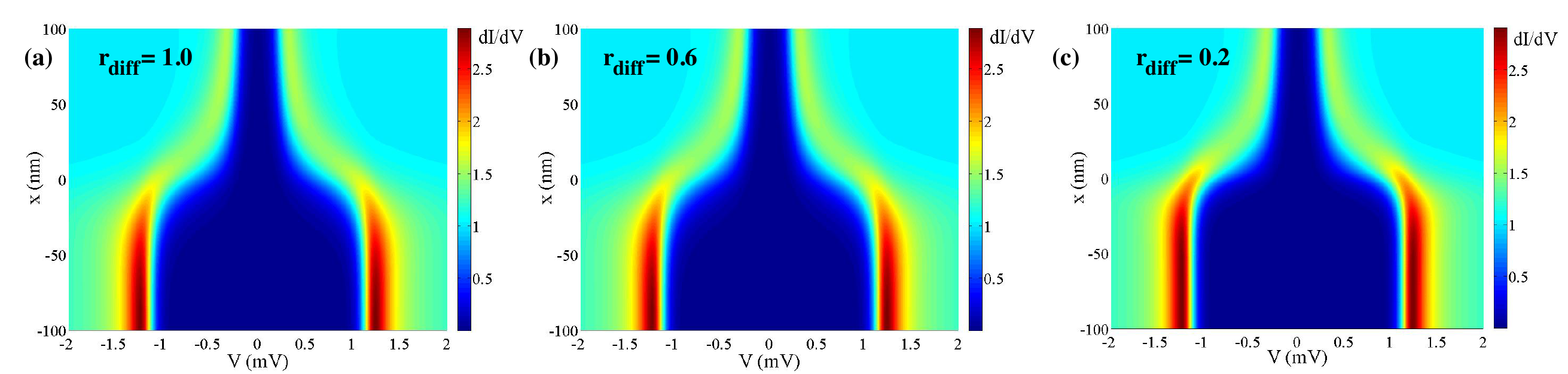} \end{center}
\caption{Normalized tunneling spectra as a function of the bias voltage and the position measured
with respect to the island edge. These results were obtained in self-consistent calculations
that take into account the inverse proximity effect in the island. The three panels correspond
to different values of the diffusion constant ratio $r_{\rm diff} = D_1/D_2$. The different parameter
values are: $T= 0.55$~K, $\Delta_1 = 1.2$~meV, $\Delta_2 = 0.23$~meV, $r=0.0$, and $D_2 = 7.3$~cm$^2$/s.}
\label{fig_IPE_thy}
\end{figure*}

In Fig.~\ref{fig_IPE_thy} we present the results of the extension of our model to account for the
inverse proximity effect. For simplicity, we have assumed a perfectly transparent interface and
the spatially-resolved spectra are shown for different values of the ratio $r_{\rm diff} =
D_1/D_2$, where $D_1$ and $D_2$ are the diffusion constants of the island and of the monolayer, respectively. 
As in the examples of the previous appendix we have used the following parameter values: $T=0.55$~K, 
$\Delta_1 = 1.2$~meV, $\Delta_2 = 0.23$~meV, and $D_2 = 7.3$~cm$^2$/s. These results suggest that the 
weak proximity effect observed inside the island could be naturally described within our model by simply 
assuming that $D_1$ is much smaller than $D_2$ (around ten times smaller). However, given the small value 
of $D_2$, it seems unlikely that the Pb island could have such a small diffusion constant. We believe that 
this result might be an artifact of our 1D model. A more natural explanation for the weak proximity effect 
would invoke the higher dimensionality of the island, which surely leads to a quick geometrical dilution of 
the order parameter inside. The confirmation of this idea would require performing 3D simulations, which are out
of the scope of this work. For this reason, and since the proximity effect inside the crystalline monolayer 
is indeed the main problem of interest in this work, we have decided to stick in the main text to the model 
in which the island is considered as a perfect superconducting reservoir.

%


\end{document}